\documentclass[usenatbib,useAMS]{mn2e}

\newcommand{\apj}{ApJ}
\newcommand{\aap}{A\&A}

\usepackage{color}
\usepackage{graphicx}
\usepackage{amsmath}
\usepackage{amssymb}
\usepackage{perpage} %the perpage package
\MakePerPage{footnote} %the perpage package command
\title[Extrapolate Dust Polarization from 350 to 150 GHz?]{Searching for Inflationary B-modes: Can dust emission properties be
  extrapolated from 350 GHz to 150 GHz?}
\author[Tassis and Pavlidou]{Konstantinos Tassis$^{1,2}$ \thanks{tassis@physics.uoc.gr} and Vasiliki Pavlidou$^{1,2}$\\
$^{1}$Department of Physics and ITCP, University of Crete, 71003
  Heraklion,Greece \\
$^{2}$Foundation for Research and Technology - Hellas,
  IESL, Voutes, 7110 Heraklion, Greece}
\begin{document}
\maketitle

\label{firstpage}

\begin{abstract}
Recent {\em Planck} results have shown that radiation from the cosmic
microwave background passes through foregrounds in which aligned dust
grains produce polarized dust emission, even in regions of the sky with the lowest
level of dust emission. One of the most commonly used ways to remove
the dust foreground is to extrapolate the polarized dust emission signal
from frequencies where it dominates (e.g., $\sim$350 GHz) to frequencies
commonly targeted by cosmic microwave background experiments (e.g.,
$\sim$150 GHz). In this paper, we describe an interstellar medium effect that
can lead to decorrelation of the dust emission polarization pattern
between different frequencies due to multiple contributions along the
line of sight. Using a simple 2-cloud model we show that there are two
conditions under which this decorrelation can be large: (a) the ratio
of polarized intensities between the two clouds changes between the
two frequencies; (b) the magnetic
fields between the two clouds contributing along a line of sight are
significantly misaligned. In such cases, 
{\em the 350 GHz polarized sky map is not predictive of that at 150 GHz}. 
We propose a possible correction for this effect, using information
from optopolarimetric surveys of dichroicly absorbed starlight. 
\end{abstract}

\begin{keywords}
cosmology: inflation -- polarization -- ISM: magnetic fields -- ISM: dust --
cosmic background radiation -- cosmology: observations 
\end{keywords}

\section{Introduction}

The recently claimed
detection, at high confidence, of B-modes in cosmic microwave
background (CMB) polarization that
cannot be attributed to lensed E-modes by the BICEP2 experiment
\citep{BICEP2} was greeted with both 
enthusiasm and caution. If all or part of
this B-mode signal were confirmed to be primordial, it would constitute
the first detection of a smoking-gun for inflation, a direct probe of
yet-unknown physics such as quantum gravity, and thus one of the
most important discoveries in astrophysics, cosmology, and high-energy
physics in the past several decades. However, a joint analysis of 
BICEP2 and {\it Planck} data \citep{jointPB} showed that most, if not all, of the
BICEP2 signal was the result of the most severe, ever-present
contaminant in the studies of CMB polarization: polarized emission
from interstellar dust. This development does not slow down the search
for primordial B-modes; but it does highlight the need for a
much-enhanced ability to understand and subtract foregrounds.

 {\it
  Planck} has now established that even in sky regions away from the
Galactic plane, where the dust emission is minimum, the primordial
B-mode signal is below that of dust  \citep{Planckang}. 
In order for a future B-mode experiment to detect a primordial signal,
the polarized emission from interstellar dust and its polarization
pattern will have to be subtracted from the CMB sky at high
accuracy. Any systematic errors entering this correction will need to
be carefully considered: even if they are small compared to the level
of the dust signal, they may be imporatnt compared to the (currently
unknown) level of the primordial signal. It is such a systematic
effect that we describe here. 

Currenty the subtraction of the dust signal is based on measuring
polarized emission from dust at frequencies where it is dominant
(e.g., $\sim$350 GHz) and extrapolating the dust polarization pattern
to frequencies dominated by the CMB and targeted by CMB
experiments (60 - 150 GHz). The
basis of this approach
is that the dust emission spectrum is reasonably well-understood
(typically represented by a modified blackbody spectrum); and thus the
dust polarization pattern measured at 350 GHz can be extrapolated to lower frequencies 
\citep{Planckang}. The polarization direction at 350 GHz is dictated by the
magnetic field threading the interstellar clouds where the dust
resides. If for example all emission originates from a single cloud,
then the emission will be partially polarized in a direction
perpendicular to the plane-of-the-sky (POS) projection of the magnetic
field. The dust emission at 150 GHz should then be
polarized in the same direction - since the polarization is generated
by that same magnetic field. 

The validity of such an extrapolation 
breaks down whenever more than one ``cloud'' (more than one dusty region
with different temperatures and magnetic fields) is contributing to the integrated signal
along the line of sight. The reason is that the total linearly polarized intensity fraction and polarization
  direction is obtained by algebraic addition of each of the Stokes linear
  polarization parameters $Q$, $U$, and $I$ along the line of sight.  If the
  relative contribution from two regions with different polarization
  directions (effectively different magnetic field directions)
  changes between frequencies, so will the resulting polarization
  fraction and polarization direction. Indeed, the relative contribution
  {\em will} change between frequencies if the two regions have
  different temperatures, due to the temperature-dependent black-body
  part of the dust 
  emission spectrum. 
%Therefore unless all of the line-of-sight signal is due to dust at the same
%temperature, or due to dust residing in magnetic fields identically
%oriented on the POS, the resulting polarization direction
%will change between frequencies. 

In this paper we use a simple two-cloud model to demonstrate that this
effect could cause decorrelation of the polarization pattern between
350GHz and 150GHz maps. We calculate conditions under which the
effect would be large in configuration (rather than in Fourier)
space. And we propose ways to correct for this systematic effect. 
% and caution against the
%extrapolation of polarized dust emission properties from 350 GHz to
%150 GHz without proper evaluation of the number of potential
%contributors along the line of sight. 

\section{Two-cloud model}\label{model}

\begin{figure}
\begin{center}
\includegraphics[width=6cm, clip=true]{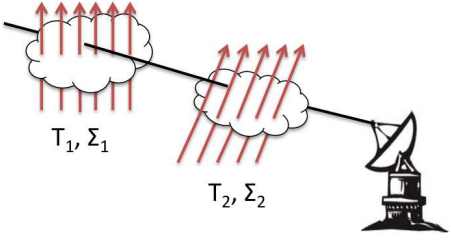}
\caption{\label{cartoon} Two clouds with temperatures $T_1$ and $T_2$,
  dust column densities $\Sigma_1$ and $\Sigma_2$, and magnetic fields
oriented at different directions contribute to the total dust-emission
intensity along a line of sight.}
\end{center}
\end{figure}

We consider a simple case in which only two ``clouds'' of temperatures $T_1$
and $T_2$ and dust column densities $\Sigma_1$ and $\Sigma_2$
contribute to the total dust emission along a line of sight (see Fig.~\ref{cartoon}). 
The total intensity emitted by each cloud can be generally well
described by a modified black-body spectrum \citep{Hil1983,planckxi, planckxxix}, 
\begin{equation}
I_\nu = B_\nu(T)\left[1-\exp(-\kappa \Sigma)\right],
\end{equation}
where $B_\nu(T)$ is the blackbody spectral radiance and $\kappa$ is
the opacity. 
%For simplicity, we will assume that the opacity as a function of
%frequency, $\kappa(\nu)$ is the same in the two clouds. Finally, we will
%assume that the clouds are partially linearly polarized, i.e. that the
%Stokes parameter $V$ is in both cases equal to zero. 

If the degree of polarization in a cloud is $p$, then the polarized
intensity will be $pI$. In this paper we assume that $p$ is
independent of frequency, although in the general case this is not
so \citep{serkowski}.
In the case of low optical depth $\kappa \Sigma$ (appropriate for
regions of low dust emission, such as the low--dust-emission regions
typically targeted by CMB experiments), the ratio of polarized 
intensities contributed by each cloud can be written as
\begin{equation}\label{eqr}
\rho_\nu = \frac{I_1}{I_2}\frac{p_1}{p_2} = \frac{B_\nu(T_1)}{B_\nu(T_2)}\frac{\Sigma_1}{\Sigma_2}\frac{p_1}{p_2}.
\end{equation}
This will be a function of frequency because of the
blackbody part of the emission spectrum. 

The polarization state of the dust emission from each cloud can be
completely described by the Stokes parameters $Q$, $U$, and $I$ (where
$I$ is the total intensity). The polarization degree $p$ and polarization
angle $\chi$ of the emission from a single cloud are related to the
Stokes parameters through 
\begin{equation}\label{parameters}
p = \frac{\sqrt{Q^2+U^2}}{I}\,,\,\,\,\tan 2\chi = \frac{U}{Q}\,.
\end{equation}
When both clouds contribute to the emission along a given line of
sight, the corresponding Stokes parameters add:
\begin{equation}\label{addition}
Q_{\rm tot} =Q_1+Q_2\,,\, U_{\rm tot} = U_1 + U_2 \,,\, I_{\rm
  tot} = I_1+I_2 \,.
\end{equation}
The polarization degree $p_{\rm tot}$ and electric vector position angle $\chi_{\rm tot}$ of the total emission then are
obtained by inserting $Q_{\rm tot}$, $U_{\rm tot}$, and $I_{\rm tot}$ into
Eq.~(\ref{parameters}). 

If  the POS-projection of the magnetic field in cloud 1 lies
in a direction forming an angle $\alpha$ with that of cloud 2, the polarization
directions of each cloud's emitted radiation will similarly form an angle $\alpha$. Without loss of
generality, we can take the magnetic field of cloud 1 to be at an
angle $\chi_1=\alpha/2$ from the reference direction, and that of cloud 2
to be at an angle $\chi_2 = -\alpha/2$. Then $\tan2\chi_1 = \tan \alpha$,
$\tan2\chi_2=-\tan\alpha$, and $p_i=|Q_i|\sqrt{1+\tan^2\alpha}/I_i$ in
both cases ($i=1,2$). 
We can represent this configuration with Stokes parameters
\begin{equation}
I_1=I_2\rho_\nu(p_2/p_1)\,, Q_1 = \frac{\rho_\nu p_2I_2}{\sqrt{1+\tan^2\alpha}}\,, U_1 =
\frac{\rho_\nu p_2I_2 \tan\alpha }{\sqrt{1+\tan^2\alpha}}
\end{equation}
and
\begin{equation}
I_2\,, Q_2 = \frac{p_2I_2}{\sqrt{1+\tan^2\alpha}}\,, U_2 =
-\frac{p_2I_2 \tan\alpha }{\sqrt{1+\tan^2\alpha}}\,,
\end{equation}
where we have taken $Q_1,Q_2 >0$, and $U_1,U_2$ carry the
signs. 
The Stokes parameters of the total signal will then be
\begin{eqnarray}
&&Q_{\rm tot} = \frac{p_2I_2(\rho_\nu+1)}{\sqrt{1+\tan^2\alpha}} \,,\,\,
U_{\rm tot} =\frac{p_2I_2 (\rho_\nu-1) \tan\alpha }{\sqrt{1+\tan^2\alpha}}\,,\,\, \nonumber \\
&&I_{\rm tot} = I_2(1+\rho_\nu p_2/p_1).
\end{eqnarray}
The resulting polarization parameters will thus be 
\begin{eqnarray}\label{general}
&& \tan2\chi_{\rm tot} = \tan\alpha\frac{(\rho_\nu-1)}{(\rho_\nu+1)} \nonumber\\
&& p_{\rm tot} = \frac{p_2}{1+\rho_\nu p_2/p_1} 
\sqrt{\frac{(\rho_\nu+1)^2+\tan^2\alpha (\rho_\nu-1)^2}{1+\tan^2\alpha}}\,.
\end{eqnarray}

We next consider the two extreme cases of relative magnetic
field directions: parallel and perpendicular. 

\subsection{Magnetic fields parallel}
The simplest and least problematic case is the one where the
directions of the POS magnetic field, and thus the
polarization directions, are parallel in the two clouds (i.e. $\alpha =0$). In this case,
Eqs.~(\ref{general}) reduce to 
\begin{equation}
\tan2\chi_{\rm tot} = 0\,,\, p_{\rm tot} = \frac{p_1p_2
  (\rho_\nu+1)}{p_1+\rho_\nu p_2}\,,
\end{equation}
which simplifies to $p_{\rm tot}=p_0$ if $p_1=p_2=p_0$. If the
polarization properties of both couds are the same, the resulting
combined emission will also share these properties. 

\subsection{Magnetic fields perpendicular}\label{perp}

Conversely, the most worrisome case is the one where the
POS-projection of the magnetic field in cloud 1 lies
in a direction perpendicular to that in cloud 2. In this case, $\tan
\alpha \rightarrow \infty$, so from Eqs.~(\ref{general}) we obtain 
\begin{equation}
\chi_{\rm tot} \approx 45^\circ{\rm sgn} (\rho_\nu-1)\,, 
p_{\rm tot} \approx \frac{p_1p_2|\rho_\nu - 1|}{p_1+\rho_\nu p_2}\,.
\end{equation}
The polarization angle 
$\tan2\chi_{\rm
tot} \rightarrow \infty$ and $\chi \rightarrow 45^\circ$ if 
cloud 1 dominates in polarized emission; conversely, $\tan2\chi_{\rm
tot} \rightarrow -\infty$ and $\chi \rightarrow -45^\circ$ if
cloud 2 dominates in polarized emission. If $\rho_\nu-1$ retains the same sign between two
frequencies, the resulting polarization angles will be the same. If
$\rho_\nu-1$ changes sign, the resulting polarization angles {\em will abruptly
rotate by $90^\circ$ as one changes from one frequency to another}.

\section{Results}\label{results}
\begin{figure}
\begin{center}
\includegraphics[width=7cm, clip=true]{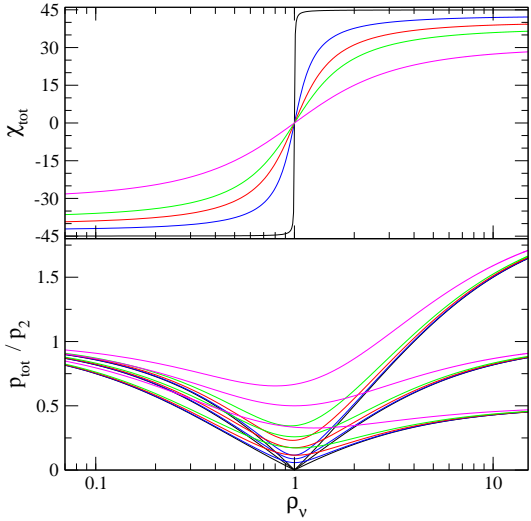}
\caption{\label{twoofrho} Upper panel: $\chi_{\rm tot}$ as a function
  of $\rho_\nu$ for different values of  angle $\alpha$
between the POS magnetic fields of the clouds: black:
$\approx 90^\circ$; blue: $85^\circ$; red: $80^\circ$; green: 75$^\circ$;
magenta:$60^\circ$. Lower panel: $p_{\rm tot}/p_2$ as a function of
$\rho_\nu$. Colours as above. Three different cases are shown, with
$p_2/p_1=0.5$, $p_2/p_1=1$, and $p_2/p_1=2$, distinguished by their
different asymptotic behavior at high $\rho_\nu$, when $p_{\rm tot
 } \rightarrow p_1$. 
}
\end{center}
\end{figure}

\begin{figure}
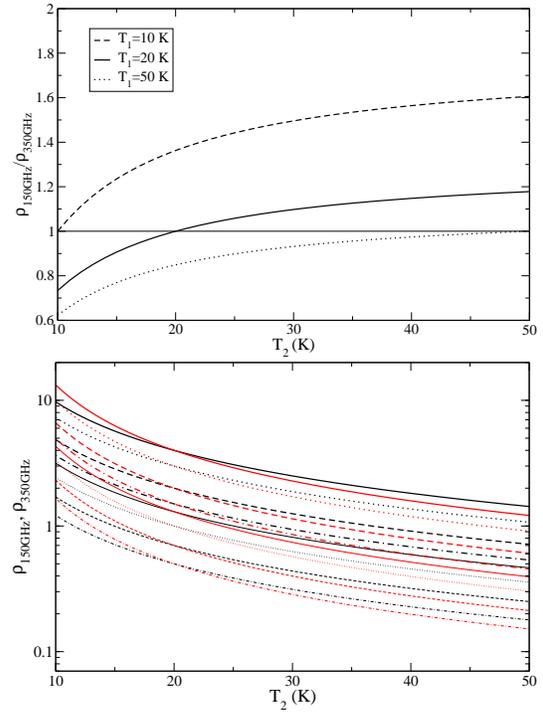

\begin{center}
\includegraphics[width=7cm, clip=true]{r350o3150.eps}
\includegraphics[width=7cm, clip=true]{rr.eps}
\caption{\label{roverr} Upper panel: ratio of polarized intensities $\rho$ at 350 GHz and 150 GHz as
  a function of cloud 2 temperature, for three different cloud 1
  temperatures. Lower panel: Ratio $\rho$ at 150 GHz (black) and 350 GHz
  (red),  as a function of cloud 2
  temperature. Cloud 1 temperature is fixed at 20 K. Line type
  corresponds to different values of the product 
  $(\Sigma_1/\Sigma_2)(p_1/p_2)$: dot-dash: 0.5; dash: 0.7; dot: 1; solid: 1.3;
  thick dot-dash: 1.5; thick dashed: 2; thick dot: 3; thick solid: 4.
}
\end{center}
\end{figure}

The behavior of $\chi_{\rm tot}$ and $p_{\rm tot}$ as a function of
the ratio of polarized intensities, $\rho_\nu$, between the two
clouds, is demonstrated in Fig.~(\ref{twoofrho}). When $\rho_\nu \ll 1$,
cloud 2 dominates, and both $\chi_{\rm tot}$ and $p_{\rm tot}$ tend to
the corresponding cloud 2 values; conversely, when $\rho_\nu \gg 1$, cloud
1 dominates. As a result,  the maximum change in $\chi_{\rm tot}$ as $\rho_{\nu}$
changes is in all cases equal to $\alpha$. This maximu change occurs when we
transition from cloud 1 completely dominating the polarized emission
to cloud 2 dominating. 
The larger the angle $\alpha$, the more abrupt the transition between the two
regimes, which occurs at $\rho_\nu = 1$. The resulting polarization fraction $p_{\rm tot}$ is plotted in
units of $p_2$, and for three values of $p_1$: $p_1=p_2/2$, $p_1=p_2$,
and $p_1=2p_2$, distinguished in the plot by the different asymptotic
behavior of the $p_{\rm tot}(\rho_\nu)$ curves as cloud 1 dominates
(for $\rho_\nu \gg 1$). The possible change in $p_{\rm tot}$ as
$\rho_\nu$ changes is larger the larger the angle $\alpha$ and the larger
the difference between $p_1$ and $p_2$. As discussed in \S
\ref{perp}, for the extreme case $\alpha \rightarrow 90^\circ$, the
resulting POS magnetic field has a step-function behavior  between
$-45^\circ$ for $\rho_\nu<1$, and $+45^\circ$ for $\rho_\nu >1$, while
the resulting polarization fraction passes through zero at
$\rho_{\nu}=1$.

In conclusion, the potential change of the observed polarization
 properties of the combined emission from the two clouds, between 150 and 350 GHz, depends on the change in $\rho_\nu$
 between the two frequencies. In turn, from Eq.~(\ref{eqr}), $\rho_\nu$ depends on the
 temperature of the two clouds, the ratio of their column densities,
 and the ratio of their polarization fractions. Note that $\rho_\nu$
 {\em does not} depend on the ratio of temperatures $T_1/T_2$, but has
 a more complicated dependence through the exponential part of the
 Planck law. In fact, when both clouds are hot enough to be in their
 Rayleigh-Jeans regime in the frequencies of interest, and the
 dependence of $\rho_{\nu}$ on the temperature {\em is} dominated by
 the $T_1/T_2$ ratio, the frequency dependence of $\rho_\nu$ cancels
 out, $\rho_\nu$ remains unchanged between 150 and 450 GHz,  and the
 polarization properties of the resulting emission remain
 constant. However, dust in typical ISM temperatures ($\sim 20$ K) is {\em
 not} in the Rayleigh-Jeans limit between 150 and 350 GHz. 
Possible values of $\rho_\nu$ for interstellar medium clouds, for
different temperatures of the two clouds and different values of the
product $(\Sigma_1/\Sigma_2)(p_1/p_2)$, as well as likely changes  of
$\rho_\nu$ between $150$ and $350$ GHz are shown in Fig.~\ref{roverr}.

\begin{figure}
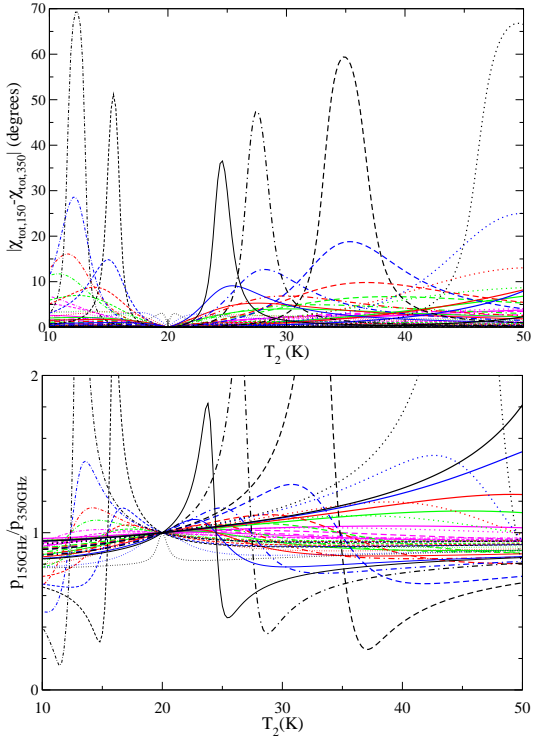

\begin{center}
\includegraphics[width=7cm, clip=true]{dchi.eps}
\includegraphics[width=7cm, clip=true]{pp.eps}
\caption{\label{pp} 
Upper panel: difference of polarization angles of combined
  signal as measured at 350 GHz and 150 GHz, as a function of cloud 2
  temperature.
Lower panel: ratio of polarization fractions of combined signal
  as measured at 350 GHz and 150GHz as a
 function of cloud 2 temperature, assuming $p_1=p_2$. Color corresponds to angle $\alpha$
between the POS magnetic fields of the clouds: black:
$\approx 90^\circ$; blue: $85^\circ$; red: $80^\circ$; green: 75$^\circ$;
magenta:$60^\circ$.  Line types as in Fig.~\ref{roverr} (lower).}
\end{center}
\end{figure}

The manner in which the polarization properties measured for the combined signal
change between 350 GHz and 150 GHz is shown in Fig.~\ref{pp}. 
Different
colours correspond to different angles $\alpha$ between the POS
magnetic fields of the two clouds as in Fig.~\ref{twoofrho}, and different line types to
different values of the product $(\Sigma_1/\Sigma_2)(p_1/p_2)$, as in Fig.~\ref{roverr}.
The
upper panel shows the difference of the polarization angle of the
combined signal as measured between 150 GHz  and 350 GHz, and the lower
panel shows the ratio $p_{\rm 150GHz}/p_{\rm 350 GHz}$, plotted in
both cases against the
temperature of cloud 2. The temperature of cloud 1 is fixed at the
average interstellar dust value of 20 K \citep{Planckfreq}. In the
case of the degree of polarization, we have taken $p_1=p_2$, and the true
effect may thus be even greater (see Fig. \ref{twoofrho}, lower panel). 

%\caption{\label{dchi} }
%\end{center}
%\end{pp}figure}
The first point to note is that once we consider the possible
polarization pattern decorrelation as a function of all parameters
entering the problem, a large range of outcomes is possible. The
physically transparent picture depicted in Fig.~(\ref{twoofrho})
becomes much more complicated. However, there are still generally
informatve trends. 

When the temperatures of the two
clouds are different, severely misaligned cloud magnetic fields will
result in significant differences in polarization degree and
polarization angle between 150 GHz and 350 GHz for a range of possible
parameters of our model. Caution is then warranted when using
350 GHz data to extrapolate dust properties to 150 GHz, as the 150 GHz
polarized emission map might look very different from the 350 GHz
map. The polarization degree might differ by large factors (more than
a factor of 2 for the most misaligned magnetic fields, more than 20\%
if the POS magnetic field of the two clouds forms an angle higher than
75$^\circ$). Similarly, the polarization angle can be up to $90^{\circ}$
different if the clouds are perfectly misaligned, and certainly higher
than 20$^\circ$ for clouds with severely but not perfectly misaligned
POS magnetic fields ($\sim 85^\circ$). 

On the other hand, there are large parts of the parameter space where
the polarization properties of the combined emission would look very
similar at the two frequencies: if the POS magnetic fields of the
two clouds form an angle of less than 60$^\circ$, then the difference
in polarization fraction is less than $10\%$ and the difference in
polarization angle less than $5^{\circ}$, independently of the other
parameters of our model. This behavior, as discussed in the next
section, offers a straghtforward way to correct for this systematic
effect, by masking problematic regions with severe magnetic field
misalignments along the line of sight.

\section{Discussion}\label{discussion}

The effect we have discussed here is intrinsic to all microwave experiments and is due to
information loss due to line-of-sight integration. Future experiments
with greater sensitivity/angular resolution will not
avoid this systematic uncertainty. However, if the magnetic
fields of the contributing clouds form an angle smaller than
$60^\circ$, then the resulting change in the polarization properties
maps from 350 GHz to 150 GHz is minimal (at least under the assumptions
of the simple model considered here). 

Our model can be improved in a variety of ways: e.g., more than two
contributing clouds along the line of sight, a physical model of
dust properties and emission, and constraining the observed combined emission
spectrum along a line of sight to agree within uncertainties with the
observed one. Additionally, in our model we assumed the degree of
polarization to be constant across
frequencies, and the frequency
dependence of the dust opacity to be identical in the two clouds. Any
deviation from these assumptions will cause further change between
frequencies in
the ratio $\rho_\nu$ of polarized emission between the two
clouds. Further change in $\rho_\nu$ will cause
further discrepancies between the maps in the
two frequencies, up to the maximum difference, which is the difference
between a state where cloud 1 completely dominates the polarized
emission, and a state where cloud 2 completely dominates. 

There are two ways for future experiments to deal with the systematic
effect we have identified here. The first is to calculate, for the specifications for
each experiment, the magnitude of the error introduced by this effect in
Fourier space, and determine whether it is comparable to the
level of the primordial signal they might claim. This would involve
simulating the effect using a detailed dust emission model, a
simulated 3-dimensional map of high-Galactic-latitude interstellar
clouds and their magnetic fields, and the specific instrument response
functions, mapping frequencies, and analysis techniques. The second is
to correct for the effect. 

Although a detailed analysis in Fourier space is beyond the scope of
this letter, some empirical data do exist addresing polarization pattern decorrelation of
the dust signal between different frequencies in Fourier space.  
Cross-correlation studies of the polarized microwave sky
between frequencies performed by \citet{Planckfreq} are limited to mid
Galactic latitudes (lower than those targeted by CMB B-mode
experiments) and average on larger (10$^\circ$) scales than those
relevant for primordial B-modes. Mid latitudes are llikely to be less
decorrelated, because a larger number of clouds is likely to be
present along each line of sight, which more frequently than not will average
out the strong variations that are possible in the case of only two
clouds. Similar studies in \citet{Planckang} at higher latitudes are limited
to a smaller frequency difference (between 353 and 217 GHz
instead of 353 abd 150). For larger frequency
differences, the decorrelation will be larger, because the
difference in polarized emission fraction from each cloud will be
larger (see Fig.~(\ref{twoofrho})). Even so, the $1\sigma$ pattern decorrelation
ratio between 353 and 217 GHz is found to be at $3\%$ (of the dominant
dust signal); assuming that it woud similarly be $\gtrsim 3\%$ at CMB B-mode
mapping frequencies, if the primordial signal is at 30\% of the dust, the
systematic discussed here would be at 10\% of the primordial signal, and if the
primordial signal is at 3\% of the dust, the systematic would be at
the 100\% level. Depending on the level of primordial
signal, this effect could be of little concern, or very problematic.

In configuration space, data are similarly not very constraining. If
we were to make the simplistic assumption that each line of sight in a
CMB B-mode experiment experiences polarized dust emission from two clouds and the
magnetic fields in these cloud are randomly (uniformly) distributed,
then the probability of a large, problematic ($>60^\circ$)
misalignment of magnetic fields is 33\%. The realistic
scenario is much more complicated: the probability of severe
misalignments depends on the number of clouds within the Galactic
disk projected to high latitudes, the number of higher-distance
high-velocity clouds that may also intervene along specific lines of
sight, the size of each of these clouds, the correlation length of the
magnetic field within a single cloud, and large-scale correlations of
the magnetic field within the Galactic disk. Certainly the sky at high
latitudes appears ``patchy'' on the plane of the sky \citep{Planckdust}. By
Copernican arguments (see also \citealp{Planckfreq,planckxx}), it will
also be patchy along the line of sight, with several distinct clouds,
each with its own, potentially different, magnetic
field. Additionally, at high latitudes, interstellar clouds are most likely to be located
nearby, from a few  tens of parsecs to one  Galaxy scale height (few
hundred parsecs), and thus span large angular scales. 
(e.g., the Polaris flare cloud, \citealp{polaris}, over $10^\circ$ 
across, Galactic latitude $b$ up to $30^\circ$,  and the Leo cloud, 
\citealp{leo}, $\sim 10^\circ$ across, $b$ up to
$50^\circ$). Halo objects can also emit in microwave frequencies
(e.g. high velocity clouds, such as  IVC 135+53, \citealp{lenz} with $b$ up to
$60^\circ$, $8^\circ$ across). Only a few severe misalignments between
such large (in angular scale) clouds may end up affecting a large fraction of the CMB
sky.  
%Information on the structure of the
%magnetic field on the $\sim 1^\circ$ scale exists (although it is
%currently sparse) from optopolarimetric surveys at high Galactic latitudes
%\citep{finnish} (stellar optopolarimetry also traces the magnetic field direction in which absorbing
%interstellar dust is embedded \citealp{roger1999}). These surveys show strong
%variations in polarization angle, and with no alignment on the degree
%scale, suggesting that the same would be true along the line of sight. 

This systematic effect could, however, be {\em corrected} through stellar optopolarimetry.
Polarization of starlight induced by dichroic dust absorption traces
the direction of the magnetic field in intervening absorbing
clouds \citep{roger1999}. If the optopolarimetric properties of many (tens) of stars
within a single CMB B-mode beam were measured, and if these stars were
located at various, known distances, then the magnetic field direction
in the various clouds along each line of sight could be
determined. Such {\em magnetic tomography} of the interstellar medium
could imeddiatey identify lines of sight with severe misalignments of
the magnetic field along the line of sight, and these lines of sight
could be simply masked from B-mode analyses, akin to the treatment of
point sources. In the era of Gaia (e.g., \citealp{gaia}) which will
provide distance measurements to stars down to 20th magnitude, the
bottleneck in such an endeavor will be the number of high-accuracy optopolarimetric
measurements that can be performed down to very low polarization
fractions characteristic of the regions targeted by CMB experiments. 

\section*{Acknowledgments}
We thank A. Readhead, B. Hensley, N.
  Kylafis, P. Goldsmith, G. Panopoulou, A. Tritsis, and I. Liodakis for useful
  comments. We thank an anonymous referee for a constructive review. We
  acknowledge support by: the ``RoboPol'' project, 
implemented under the ``ARISTEIA'' Action of the ``OPERATIONAL PROGRAMME EDUCATION AND
LIFELONG LEARNING'', co-funded by the European Social Fund (ESF) and Greek National
Resources; the European Commission Seventh Framework Programme (FP7) through 
grants: PCIG10-GA-2011-304001 ``JetPop''; PCIG-GA-2011-293531
``SFOnset''; PIRSES-GA-2012-31578 ``EuroCal''.

%\bibliography{pitfalls}

\end{document}